\documentclass{PoS}
\usepackage{multirow,axodraw}
\newcommand{\amuhad}{{a_\mu^{\rm had}}}

\newcommand{\bi}{\begin{itemize}}
\newcommand{\ei}{\end{itemize}}
\newcommand{\unit}{\mathbf{1}}

\newcommand{\SUtwo}{{SU(2)}}
\title{New ideas for $g-2$ on the lattice}
\ShortTitle{New ideas for $g-2$}

\author{\speaker{Andreas J\"uttner}, {Michele Della Morte}\\
        Institut f\"ur Kernphysik\\
	Johannes-Gutenberg Universit\"at Mainz\\
	Johann-Joachim-Becher Weg 45\\
	55099 Mainz\\
        E-mail: \email{juettner@kph.uni-mainz.de},
        \email{morte@kph.uni-mainz.de}}

\abstract{
\vspace{-8cm}
Using iso-spin symmetry
the quark-connected and disconnected contributions to the hadronic vacuum
polarisation in a  theory with $N_f=2$ flavours can be described as independent 
correlation functions, respectively. We show how this allows to 
use twisted boundary conditions for the connected 
contribution in order to improve the $q^2$-resolution in lattice QCD.
Furthermore we derive
an exact relation between the connected and the disconnected contributions
at NLO in chiral perturbation theory.
We discuss extensions to theories with more than 2 dynamical flavours. 
\vspace{-13cm}{\begin{flushright} MKPH-T-09-23\end{flushright}}
}

\FullConference{The XXVII International Symposium on Lattice Field Theory - LAT2009\\
		 July 26-31 2009\\
		 Peking University, Beijing, China}

\begin{document}

\section{Introduction}
As recently
summarized in \cite{Jegerlehner:2009ry} the current experimental
value for the full muon anomalous magnetic moment is 
$a_\mu=11659{208(6.3)}\times 10^{-10}$
which has to be compared to the Standard Model prediction of 
$a_\mu = 11659{179(6.5)} \times10^{-10}$. Although there are discussions
about how to correctly estimate the 
size of the systematic errors there persists a 
tension between experiment and theory  which amounts to around 
three standard deviations (3.2$\sigma$ in the comparison quoted above).
The muon anomalous moment is a remarkable observable in that all
three sectors of the SM contribute considerably. While perturbation theory
is employed in order to predict QED and weak contributions, the
leading and next-to-leading hadronic contributions are 
non-perturbative effects. Lacking reliable and precise theory
computations from first principles the current SM predictions for the leading
hadronic contribution $\amuhad$ are derived
from experimental measurements of $e^+e^-$-annihilation into hadrons leading
to a world average \cite{Jegerlehner:2009ry} 
for the hadronic contribution of $\amuhad = 690.3(5.3)\times 10^{-10}$.

Table \ref{tab:prev_lat} summarizes previous attempts to compute 
$\amuhad$ in lattice QCD in comparison with the determination 
from $e^+e^-$ annihilation.
\begin{table}\label{tab:prev_lat}
\begin{center}
\begin{tabular}{c|cc}
\hline\hline\\[-4mm]
$e^+e^-$-annihilation 	&$690.3(5.3)\times 10^{-10}$&\cite{Jegerlehner:2009ry}\\
\hline\\[-4mm]
\multirow{3}{*}{LQCD}	&$713(15)\times 10^{-10}$&\multirow{2}{*}{\cite{Aubin:2006xv}}\\	
			&$748(21)\times 10^{-10}$&\\
			&$446(23)\times 10^{-10}$&\cite{Gockeler:2003cw}\\
			&in progress	 &\cite{Renner:2009by}\\
\hline\hline
\end{tabular}
\end{center}\caption{Some recent results for the leading hadronic
contribution to the muon anomalous moment, $\amuhad$.}
\end{table}
Without going into the technical details of the various attempts to compute
$\amuhad$ on the lattice it is obvious that currently the purely theoretical 
predictions can't match the level of precision 
which can be reached via the experimental determination. But yet, an independent
confirmation of the result from $e^+e^-$-annihilation matching it in precision
and by means of a SM calculation is clearly desireable.

The leading hadronic contribution to the muon anomalous moment is the
convolution integral
\begin{equation}\label{eqn:amu}
 \amuhad = 4\pi^2\left(\frac{\alpha_{EM}}{\pi}\right)^2
	\int\limits_0^\infty dK^2 f(K^2)\left(
		\Pi(K^2)-\Pi(0)\right)\,,
\end{equation}
where $K$ is the Euclidean momentum and
the function $f(K^2)$ in~(\ref{eqn:amu}) diverges for $K^2\to 0$
(see e.g. the discussion in \cite{Blum:2002ii}).
$\Pi(K^2)$ is the vacuum polarization which for a theory
with $N_f$ quark flavours is defined through
\begin{equation}\label{eqn:Pimunu}
 \Pi_{\mu\nu}^{(N_f)}(q)=(q_\mu q_\nu-g_{\mu\nu}q^2)\Pi (q^2)
=i\int d^4x e^{iqx}\langle j_\mu^{(N_f,\, EM)}(x) j_\nu^{(N_f,\, EM)}(0)
	\rangle\,,
\end{equation}
where $j_\mu^{(N_f,\,EM)}$ is the corresponding electromagnetic current.

Apart from the usual sources of systematic errors in simulations of 
lattice QCD (unphysically heavy quark mass, finite volume, cut-off effects)
we identify three sources of systematic errors that
have to be brought under control in order to be able to make
precise predictions for $\amuhad$ from first principles:
\bi
 \item[1)] since $f(K^2)$ in~(\ref{eqn:amu}) diverges for $K^2\to 0$
	it is crucial to have a good momentum resolution for $\Pi$ for
	small values of $K$. Given that the smallest Fourier mode 
	in eqn.~(\ref{eqn:Pimunu}) on the lattice
	for typical simulation parameters is $\frac {2\pi}{2L}\approx 250$MeV
        ($L$ being the spatial extent of the lattice and usually the time-extent
	is chosen as $T=2L$) one currently 
	extrapolates results for $\Pi(q^2)$ from the region 
        where lattice data is available using some phenomenological ansatz
 \item[2)] carrying out the Wick contraction in (\ref{eqn:Pimunu})
 	reveals contributions 
	from quark-disconnected diagrams which are notoriously hard 
	to compute in lattice QCD
 \item[3)] Aubin and Blum \cite{Aubin:2006xv} found strong signs for
        vector dominance in their lattice data which has to be
        modelled in order to allow for a reliable description of the 
        data within an effective theory frame work.
\ei
Here we suggest a computational strategy that should allow to systematically
reduce the uncertainties due to 1) and 2). Concerning 3) we will follow
\cite{Aubin:2006xv} as long as the simulated quark masses are unphysically 
large. They used a model where a vector particle is added to the 	
chiral Lagrangian as an additional degree of freedom 
in order to better describe their lattice data for $\Pi(q^2)$.
This work is part of a major effort by the Mainz group aiming at a
precision computation of the leading hadronic contribution to the
muon anomalous moment (see also Hartmut Wittig's talk \cite{Wittig:2009lat}).

\section{New strategy for $N_f=2$}
Partially twisted boundary conditions 
\cite{Bedaque:2004kc,deDivitiis:2004kq,Sachrajda:2004mi, Bedaque:2004ax,Tiburzi:2005hg,Flynn:2005in,Guadagnoli:2005be,Tiburzi:2006px,Bunton:2006va}
have now been used successfully
to improve computations of observables of processes that depend on the
hadron momentum. However, as already stated in \cite{Sachrajda:2004mi}
the net effect of the twist in flavour neutral hadrons, like e.g. the $\pi^0$
 is zero since the twist of the
quark and the anti-quark cancel. The situation is similar here - the
electromagnetic current is flavour diagonal, hence a naive application of
(partially) twisted boundary conditions is ruled out.

The correlator entering eqn.~(\ref{eqn:Pimunu}) for $N_f=2$ is
\begin{equation}
 \langle j_\mu^{(2,EM)}j_\nu^{(2,EM)} \rangle =
  \frac 49 \langle j_\mu^{uu}j_\nu^{uu}\rangle
 -\frac 29 \langle j_\mu^{uu}j_\nu^{dd}\rangle
 -\frac 29 \langle j_\mu^{dd}j_\nu^{uu}\rangle
 +\frac 19 \langle j_\mu^{dd}j_\nu^{dd}\rangle\,,
\end{equation}
where we make the flavor content of the quark bilinear currents explicit 
and where we have factorized out the electromagnetic charges of the quarks 
in units of $e$ on the r.h.s.. The first
and the last correlator on the r.h.s. 
receive contributions from both a quark-connected and 
a disconnected diagram and the second and third correlator are disconnected, 
only. After carrying out the Wick 
contractions and   using iso-spin symmetry
($m_q\equiv m_u=m_d$), 
the two point function one has to compute in a lattice simulation is
\begin{equation}
\begin{array}{rcl}
C^{(2,EM)}_{\mu\nu}(q)&=&
        \sum\limits_{x}e^{i q x}\Big\{
        \Big<\frac 59
        {\rm Tr}\left\{
        {\bar q(x)}\gamma_\nu {q(x)}
        {\bar q(0)}\gamma_\mu {q(0)}
        \right\}+\\
        &&
        \qquad\;\;\;\;\;\,\,\,\frac 19{\rm Tr}\big\{
        \gamma_\nu {q(x)\bar q(x)}
        \big\}
        {\rm Tr}\big\{
        \gamma_\mu {q(0)\bar q(0)}
        \big\}\Big>\Big\}\,,\\[2mm]
\end{array}
\end{equation}
which again using iso-spin can be written in the equivalent form
\begin{equation}
\begin{array}{rcl}
C^{(2,EM)}_{\mu\nu}(q)&=&
        { \sum\limits_{ x}e^{iqx}}\Big\{
        \frac 59
        \underbrace{\Big<{\rm Tr}\left\{
        { \bar u(x)}\gamma_\nu { d(x)}
        { \bar d(0)}\gamma_\mu { u(0)}
        \right\}\Big>}_{\equiv C^{(2,{ \rm con})}_{\mu\nu}(q)}+\\[3mm]
        &&
        \qquad\;\;\,\,\,\frac 19 
        \underbrace{\Big<{\rm Tr}\left\{
        \gamma_\nu{ d(x)\bar d(x)}
        \right\}
        {\rm Tr}\left\{
        \gamma_\mu { u(0)\bar u(0)}
        \right\}\Big>}_{\equiv C^{(2,{ \rm disc})}_{\mu\nu}(q)}\Big\}\,.\\[2mm]
\end{array}
\end{equation}
By using iso-spin we are able to express the vacuum polarization in terms of 
two  correlation functions  $C^{(2,{ \rm con})}_{\mu\nu}(q)$
and $C^{(2,{ \rm disc})}_{\mu\nu}(q)$ with their individual continuum 
and infinite volume limits. The crucial point to note is that in this way
the coupling
of the photon to the quark in the connected piece has been 
replaced by a coupling to a flavour non-diagonal current  and partially
twisted boundary conditions can be applied.
We note that a similar trick has been used for computations of the 
electromagnetic pion form factor in \cite{Jiang:2006gna,Boyle:2007wg}.

We are still left with the question of how to treat the disconnected part. To
this end we resort to the description of the vacuum polarization in $SU(2)$
chiral perturbation theory at NLO. In $SU(2)$ the electromagnetic current
receives the following singlet and non-singlet contributions,
\begin{equation}
 \begin{array}{rcc@{\hspace{5mm}}rcc}
  j_\mu^{uu}&=&\frac 12\bar\psi ({ \sqrt{2}\sigma_0}
                +\sigma_3)\gamma_\mu \psi\,,&
  j_\mu^{dd}&=&\frac 12\bar\psi ({ \sqrt{2}\sigma_0}
                -\sigma_3)\gamma_\mu \psi\,,
        \\[1mm]
  j_\mu^{ud}&=&\frac 12\bar\psi ({ \sigma_1}
                +i\sigma_2)\gamma_\mu \psi\,,&
  j_\mu^{du}&=&\frac 12\bar\psi ({ \sigma_1}
                -i\sigma_2)\gamma_\mu \psi\,,
        \\[1mm]
 \end{array}
\end{equation}
where $\psi^T=(u,d)$, $\sigma_i$ for $i=1,2,3$ are the
Pauli matrices and $\sigma_0=\sqrt{1/2}\,\unit_{2\times 2}$. The effective
theory that describes the $N_f=2$ vacuum polarization is $SU(2)$ 
chiral perturbation theory as formulated in \cite{Gasser:1983yg} 
but including at $O(p^4)$ terms with non-vanishing flavour-trace,
\begin{equation}
        \mathcal{L}^{(4)}=
        \underbrace{2l_5\langle U^\dagger \hat L_{\mu\nu}U\hat R^{\mu\nu}\rangle+
        4h_2\langle 
          \hat R_{\mu\nu}\hat R^{\mu\nu}+\hat L_{\mu\nu}\hat L^{\mu\nu}\rangle
        }_{\rm flavour\; off-diagonal}
        +\underbrace{ 4h_s\langle  R_{\mu\nu}+ L_{\mu\nu} \rangle
                 \langle  R^{\mu\nu}+ L^{\mu\nu} \rangle}_{
        \rm flavour-diagonal}\,,
\end{equation}
where angular brackets indicate a flavour trace and 
the \emph{hat} indicates that the latter has been subtracted (cf. 
\cite{Golowich:1995kd} for further details on the notation). Note the
additional low energy constant $h_s$ multiplying the flavour diagonal
term.

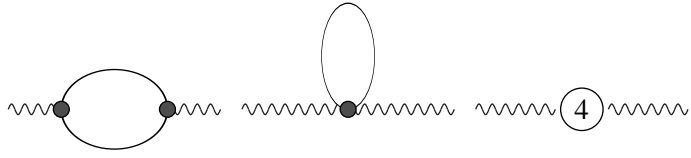
\begin{figure}
\begin{center}
    \begin{minipage}{.6\linewidth}
    \begin{picture}(80,40)(-40,-11)
        \SetScale{1}
        \Photon(-40,0)(-20,0){2}{4}
        \Photon(20,0)(40,0){2}{4}
        \GOval(0,0)(15,20)(0){1}
        \GCirc(-20,0){3}{.3}
        \GCirc(20,0){3}{.3}
    \end{picture}\hspace{2mm}
    \begin{picture}(80,40)(-40,-11)
        \SetScale{1}
        \Photon(-40,0)(40,0){2}{16}
        \GOval(0,20)(20,10)(00){1}
        \GCirc(0,0){3}{.3}
    \end{picture}\hspace{2mm}
    \begin{picture}(80,40)(-40,-11)
        \SetScale{1}
        \Photon(-40,0)(-10,0){2}{6}
        \Photon(10,0)(40,0){2}{6}
        \GCirc(0,0){8}{1}
        \Text(0,0)[c]{4}
    \end{picture}
        \end{minipage}\\[2mm]
\end{center}\caption{Diagrams in chiral perturbation theory that contribute
 to the vacuum polarization at NLO.}\label{fig:VP_NLO_diags}
\end{figure}

The diagrams contributing at NLO are illustrated in figure 
\ref{fig:VP_NLO_diags} and our preliminary result is
\begin{equation}\label{eqn:resSU2}
\begin{array}{r@{\hspace{0mm}}c@{\hspace{1mm}}l@{\hspace{1mm}}l@{
                \hspace{1mm}}l}
	 \Pi_{\mu\nu}^{(2)}(q)& = &
   \Pi_{\mu\nu}^{(3,3)}(q)+
        { \frac 19\Pi_{\mu\nu}^{(0,0)}(q)}\\[2mm]
        &=&-(q_\mu q_\nu-g_{\mu\nu}q^2)\left(
 i4\bar B_{21}(q^2,m^2_\pi)+2l_5(\mu_0)+4h_2(\mu_0)+\frac 4{ 9} 
        { h_s}+\frac{1}{48\pi^2}\ln\left(\frac {m_\pi^2}{\mu_0^2}\right)\right)\,,
        \nonumber
\end{array}
\end{equation}
where $\mu_0$ is the renormalization scale
and  $\bar B_{21}(q^2,m^2_\pi)$ is a loop integral 
	(cf. \cite{Golowich:1995kd}). 
The contributions $\Pi_{\mu\nu}^{(a,a)}$ ($a=0,1,2,3$) are the effective theory
descriptions of 2pt-correlators constructed of the currents 
$c \bar \psi \gamma_{\mu/\nu} \sigma_a \psi$, where $c\in \mathbb{C}$ is a normalization.
We add that 
$\Pi^{(0,0)}(q^2)=\Pi^{(0,0)}$ is momentum independent
at NLO and only depends on the low energy constant $h_s$. 

Alternatively it is possible to express the quark-connected and disconnected
contributions individually in the effective theory (the linear 
combination of which should of course again yield eqn.~(\ref{eqn:resSU2})).
We computed the corresponding expressions as
\begin{equation}\label{eqn:resSU2_2}
  \begin{array}{r@{\hspace{0mm}}c@{\hspace{1mm}}l@{\hspace{1mm}}c@{
          \hspace{1mm}}l}
  \Pi^{(2, \rm con)}(q^2) &=& 
     \frac {10}9\Pi^{(3,3)}(q^2) \,,\\[3mm]
   \Pi^{(2, \rm disc)}(q^2) &= &
     \frac 19\left({\Pi^{(0,0)}}-
          \Pi^{(3,3)}(q^2)\right)\,.
  \nonumber
 \end{array}
\end{equation}
Since $\Pi^{(0,0)}$ is independent of $q^2$, the quark-disconnected 
part turns out to have the same momentum dependence as the connected part
up to a finite shift proportional to $h_s$, 
\begin{equation}
 \begin{array}{rcl}
  \Pi^{(2,\rm disc)}(q^2)&=&\frac 19 \Pi^{(0,0)}-
                \frac {1}{10}\Pi^{(2,\rm con)}(q^2)\,.\\[2mm]
 \end{array}
\end{equation}
Assuming the validity of chiral perturbation theory at NLO the knowledge of
$h_s$ allows to fully predict the quark-disconnected contribution. 

The leading hadronic contribution to the muon anomalous moment
is computed from 
\begin{equation}
\begin{array}{rcl}
 \hat\Pi(q^2)&\equiv&\Pi(q^2)-\Pi(0)\,.
\end{array}
\end{equation}
One immediate observation is, that at NLO in chiral perturbation theory
this expression is free of low energy constants since the loop integrals
carry all the momentum dependence. A crucial consequence
which to our knowledge hasn't been realized before is, that e.g. in
the case of $\SUtwo$,
\begin{equation}\label{eqn:disc_prediction}
 \frac{\hat\Pi(q^2)|_{\rm disc}}{ \hat\Pi(q^2)|_{\rm con}}=
	-\frac 1{10}\,.
\end{equation}
This is a remarkable result which tells us that at this order
of the effective theory the
disconnected part shifts the central value of the connected part by
minus ten per cent.

\section{Conclusions and outlook}
To summarize, 
we have shown how partially twisted boundary conditions can be used to 
improve the momentum resolution in lattice computations of the 
hadronic vacuum polarization of a photon. 
It was shown how to analytically predict at NLO in chiral perturbation 
theory the contribution of quark-disconnected diagrams to the leading
hadronic contribution of the muon anomalous magnetic moment. It turns out
that it's effect is to reduce the contribution of the connected part by 10\%. 

We are currently working on extending these arguments to the case of 
$N_f=2+1$ flavours. One complication there is that the iso-spin argument
that allowed us to write the  quark-connected and disconnected pieces
as individual correlation functions in the case of $N_f=2$
does not work straight forwardly since
the strange quark doesn't have an iso-spin partner. Naively the 
$\bar s\gamma_\mu s$-contribution to the EM current can therefore not be
written in terms of a flavor off-diagonal current. Our  strategy
is to extend the flavour group from $SU(3)$ to $SU(3+1|1)$, i.e. to 
a graded flavour group with an additional partially quenched quark 
(which we call $r$-quark) that behaves like a mass-degenerate \emph{iso-partner}
of the $s$-quark.
Then the quark-connected
and disconnected pieces can be written in terms of individual 
correlation functions.

Since quark-disconnected  diagrams are extremely difficult to
compute on the lattice, we will investigate if our method to predict
their contribution in the frame work of chiral perturbation theory
can be applied to other phenomenologically interesting observables 
in QCD.

Following \cite{Sachrajda:2004mi} we are also computing the finite volume 
effects for the vacuum polarization including the effect of partial twisting.

Another task that remains to be finished is the inclusion of vector degrees
of freedom in order to be able to  extrapolate the lattice data
to physical quark masses and to assess the vector particle's
 effect on the stability of eqn.~(\ref{eqn:disc_prediction}).

As a joint effort the Mainz group  has 
implemented correlation functions relevant for a lattice computation of
$\amuhad$ in a C-code.
First computations  on gauge configurations
of $N_f=2$ non-perturbatively improved Wilson fermions which were generated  
as a collaborative effort within CLS (cf. the talks by 
Hartmut Wittig's \cite{Wittig:2009lat}
and Stefan Sch\"afer \cite{Bar:2009qa} at this conference) have been carried
out on the Mainz Wilson-Cluster.

{\bf Acknowledgements:} We warmly thank Dalibor Djukanovic, Jambul Gegelia,
Chris Sachrajda, Stefan Scherer and Hartmut Wittig for valuable discussions.

\bibliographystyle{JHEP}
\bibliography{g-2}

\providecommand{\href}[2]{#2}\begingroup\raggedright\begin{thebibliography}{10}

\bibitem{Jegerlehner:2009ry}
F.~Jegerlehner and A.~Nyffeler, {\it {The Muon g-2}},  {\em Phys. Rept.} {\bf
  477} (2009) 1--110, [\href{http://xxx.lanl.gov/abs/0902.3360}{{\tt
  0902.3360}}].

\bibitem{Aubin:2006xv}
C.~Aubin and T.~Blum, {\it {Calculating the hadronic vacuum polarization and
  leading hadronic contribution to the muon anomalous magnetic moment with
  improved staggered quarks}},  {\em Phys. Rev.} {\bf D75} (2007) 114502,
  [\href{http://xxx.lanl.gov/abs/hep-lat/0608011}{{\tt hep-lat/0608011}}].

\bibitem{Gockeler:2003cw}
{\bf QCDSF} Collaboration, M.~G{\"o}ckeler {\em et.~al.}, {\it {Vacuum
  polarisation and hadronic contribution to muon g-2 from lattice QCD}},  {\em
  Nucl. Phys.} {\bf B688} (2004) 135--164,
  [\href{http://xxx.lanl.gov/abs/hep-lat/0312032}{{\tt hep-lat/0312032}}].

\bibitem{Renner:2009by}
D.~B. Renner and X.~Feng, {\it {Hadronic contribution to g-2 from twisted mass
  fermions}},  \href{http://xxx.lanl.gov/abs/0902.2796}{{\tt 0902.2796}}.

\bibitem{Blum:2002ii}
T.~Blum, {\it {Lattice calculation of the lowest order hadronic contribution to
  the muon anomalous magnetic moment. }},  {\em Phys. Rev. Lett.} {\bf 91}
  (2003) 052001, [\href{http://xxx.lanl.gov/abs/hep-lat/0212018}{{\tt
  hep-lat/0212018}}].

\bibitem{Wittig:2009lat}
H.~Wittig, {\it {Mesonic and Baryonic correlation functions at fine lattice
  spacings {\rm (Talk at this conference)}}}, .

\bibitem{Bedaque:2004kc}
P.~F. Bedaque, {\it Aharonov-{B}ohm effect and nucleon nucleon phase shifts on
  the lattice},  {\em Phys. Lett.} {\bf B593} (2004) 82--88,
  [\href{http://xxx.lanl.gov/abs/nucl-th/0402051}{{\tt nucl-th/0402051}}].

\bibitem{deDivitiis:2004kq}
G.~M. de~Divitiis, R.~Petronzio, and N.~Tantalo, {\it On the discretization of
  physical momenta in lattice {QCD}},  {\em Phys. Lett.} {\bf B595} (2004)
  408--413, [\href{http://xxx.lanl.gov/abs/hep-lat/0405002}{{\tt
  hep-lat/0405002}}].

\bibitem{Sachrajda:2004mi}
C.~T. Sachrajda and G.~Villadoro, {\it {Twisted boundary conditions in lattice
  simulations}},  {\em Phys. Lett.} {\bf B609} (2005) 73--85,
  [\href{http://xxx.lanl.gov/abs/hep-lat/0411033}{{\tt hep-lat/0411033}}].

\bibitem{Bedaque:2004ax}
P.~F. Bedaque and J.-W. Chen, {\it Twisted valence quarks and hadron
  interactions on the lattice},  {\em Phys. Lett.} {\bf B616} (2005) 208--214,
  [\href{http://xxx.lanl.gov/abs/hep-lat/0412023}{{\tt hep-lat/0412023}}].

\bibitem{Tiburzi:2005hg}
B.~C. Tiburzi, {\it Twisted quarks and the nucleon axial current},  {\em Phys.
  Lett.} {\bf B617} (2005) 40--48,
  [\href{http://xxx.lanl.gov/abs/hep-lat/0504002}{{\tt hep-lat/0504002}}].

\bibitem{Flynn:2005in}
{\bf UKQCD} Collaboration, J.~M. Flynn, A.~J{\"u}ttner, and C.~T. Sachrajda,
  {\it A numerical study of partially twisted boundary conditions},  {\em Phys.
  Lett.} {\bf B632} (2006) 313--318,
  [\href{http://xxx.lanl.gov/abs/hep-lat/0506016}{{\tt hep-lat/0506016}}].

\bibitem{Guadagnoli:2005be}
D.~Guadagnoli, F.~Mescia, and S.~Simula, {\it Lattice study of semileptonic
  form factors with twisted boundary conditions},  {\em Phys. Rev.} {\bf D73}
  (2006) 114504, [\href{http://xxx.lanl.gov/abs/hep-lat/0512020}{{\tt
  hep-lat/0512020}}].

\bibitem{Tiburzi:2006px}
B.~C. Tiburzi, {\it Flavor twisted boundary conditions and isovector form
  factors},  {\em Phys. Lett.} {\bf B641} (2006) 342--349,
  [\href{http://xxx.lanl.gov/abs/hep-lat/0607019}{{\tt hep-lat/0607019}}].

\bibitem{Bunton:2006va}
T.~B. Bunton, F.~J. Jiang, and B.~C. Tiburzi, {\it Extrapolations of lattice
  meson form factors},  {\em Phys. Rev.} {\bf D74} (2006) 034514,
  [\href{http://xxx.lanl.gov/abs/hep-lat/0607001}{{\tt hep-lat/0607001}}].

\bibitem{Jiang:2006gna}
F.~J. Jiang and B.~C. Tiburzi, {\it {Flavor Twisted Boundary Conditions, Pion
  Momentum, and the Pion Electromagnetic Form Factor}},  {\em Phys. Lett.} {\bf
  B645} (2007) 314--321, [\href{http://xxx.lanl.gov/abs/hep-lat/0610103}{{\tt
  hep-lat/0610103}}].

\bibitem{Boyle:2007wg}
P.~A. Boyle, J.~M. Flynn, A.~J{\"u}ttner, C.~T. Sachrajda, and J.~M. Zanotti,
  {\it {Hadronic form factors in lattice QCD at small and vanishing momentum
  transfer}},  {\em JHEP} {\bf 05} (2007) 016,
  [\href{http://xxx.lanl.gov/abs/hep-lat/0703005}{{\tt hep-lat/0703005}}].

\bibitem{Gasser:1983yg}
J.~Gasser and H.~Leutwyler, {\it {Chiral Perturbation Theory to One Loop}},
  {\em Ann. Phys.} {\bf 158} (1984) 142.

\bibitem{Golowich:1995kd}
E.~Golowich and J.~Kambor, {\it {Two loop analysis of vector current
  propagators in chiral perturbation theory}},  {\em Nucl. Phys.} {\bf B447}
  (1995) 373--404, [\href{http://xxx.lanl.gov/abs/hep-ph/9501318}{{\tt
  hep-ph/9501318}}].

\bibitem{Bar:2009qa}
S.~Sch{\"a}fer, R.~Sommer, and F.~Virotta, {\it {Investigating the critical
  slowing down of QCD simulations {\rm (Talk at this conference)}}},
  \href{http://xxx.lanl.gov/abs/0910.1465}{{\tt 0910.1465}}.

\end{thebibliography}\endgroup
\end{document}